\documentclass[%
 reprint,
nofootinbib,
 amsmath,amssymb,
 aps,
]{revtex4-1}

\usepackage[dvipdfmx]{graphicx}
\usepackage[dvipdfmx]{}

\usepackage{dcolumn}
\usepackage{bm}

\usepackage{multirow}

\usepackage{here}
\usepackage{comment}

\usepackage{ulem}

\begin{document}

\title{Single heavy baryons with chiral partner structure in a three-flavor chiral model }

\author{Yohei Kawakami} \email{kawakami@hken.phys.nagoya-u.ac.jp}
 \affiliation{Department of Physics, Nagoya University, Nagoya, 464-8602, Japan}
 \author{Masayasu Harada}\email{harada@hken.phys.nagoya-u.ac.jp}
 \affiliation{Department of Physics, Nagoya University, Nagoya, 464-8602, Japan}

\date{\today}

\begin{abstract}
We construct an effective hadronic model including single heavy baryons (SHBs) 
belonging to the $(\mathbf{3},\mathbf{3})$ representation under $\mbox{SU}(3)_L \times \mbox{SU}(3)_R$ symmetry,
respecting the chiral symmetry and heavy-qaurk spin-flavor symmetry. 
When the chiral symmetry is spontaneously broken, the SHBs are divided into the baryons 
with negative parity of $\bar{\mathbf  3}$ representation under $\mbox{SU}(3)$ flavor symmetry
which is the chiral partners to the ones with positive parity of ${\mathbf 6}$ representation.
We determine the model parameters from the available experimental data for the masses and strong decay widths of $\Sigma_c^{(\ast)}$, $\Lambda_c (2595)$, $\Xi_c (2790)$, and $\Xi_c (2815)$. Then, we predict the masses and strong decay widths of other baryons including $\Xi_b$ with negative parity.
We also study radiative decays of SHBs including $\Omega_c^\ast$ and $\Omega_b^\ast$ with positive parity.
\end{abstract}

\maketitle

\section{\label{sec:level1}Introduction}

The spontaneous chiral symmetry breaking, which is one of the most essential properties of QCD, is expected to generate a part of hadron masses and 
causes the mass difference between chiral partners.
Investigation of chiral partner structure will provide some clues to understand the chiral symmetry.
In particular, 
study of the chiral partner structure of 
hadrons including heavy quarks gives information which are not obtained from the hadrons including only light quarks.

There are several studies of hadrons including heavy quarks based on the chiral partner structure.
The chiral partner structure of heavy-light mesons is studied in e.g., Refs.~\cite{Nowak:1992um,Nowak:1993vc,Bardeen:1993ae,Bardeen:2003kt,Nowak:2003ra}, that of doubly heavy baryons is in e.g., Refs.~\cite{Ma:2015lba,Ma:2015cfa,Ma:2017nik}, and the single heavy baryons (SHBs) are studied in e.g., Refs.~\cite{Nowak:2004jg,Harada:2012dm,Liu:2011xc,Kawakami:2018olq,Arifi:2018yhr}.

In Ref.~Ref.~\cite{Kawakami:2018olq}, we proposed a new chiral partner structure for SHBs
including a heavy quark and two light quarks.
There, we considered 
the chiral partners of $\Sigma_{Q}$ ($Q=c,b$) baryons with positive parity as $\Lambda_Q$ baryons with negative parity:
a heavy quark doublet of $\left( \Lambda_c(2595;J^P = 1/2^-) \,,\, \Lambda_c(2625;3/2^-) \right)$ is regarded as the chiral partners to the doublet of 
$\left( \Sigma_c(2455;1/2^+) \,,\, \Sigma_c(2520;3/2^+) \right)$, and $\left( \Lambda_b(5912;1/2^-) \,,\, \Lambda_b(5920;3/2^-) \right)$ to $\left( \Sigma_b(1/2^+) \,,\, \Sigma_b^\ast(3/2^+) \right)$.
Based on this structure, we predicted the pionic and photonic decay widths of these excited SHBs. The results show that,
although the decay of $\Lambda_c(2595)$ is dominated by the resonant
contribution through $\Sigma_c(2455)$, nonresonant contributions are important for $\Lambda_c(2625)$, $\Lambda_b(5912)$, and $\Lambda_b(5920)$, which reflects the chiral partner structure.
In Ref.~\cite{Arifi:2018yhr}, its experimental verification was proposed.

In the present work, we extend the chiral partner model in Ref.~\cite{Kawakami:2018olq}, which is based on $\mbox{SU}(2)_{L}\times \mbox{SU}(2)_{R}$ symmetry, to $\mbox{SU}(3)_{L}\times \mbox{SU}(3)_{R}$ symmetry. 
We consider the SHBs with negative parity which belongs to $\bm{\bar{3}}$ representation under $\mbox{SU}(3)$ flavor symmetry are the chiral partner to the SHBs with positive parity to $\bm{6}$ representation.
We introduce a chiral field belonging to $(\bm{3},\bm{3})$ representation under $\mbox{SU}(3)_{L}\times \mbox{SU}(3)_{R}$ symmetry
to construct an effective Lagrangian including the interactions to light
pseudo-scalar mesons.
Determining the model parameters from existing experimental data, we give predictions on the masses and pionic decay widths which are not experimentally determined.
We also study radiative decays by introducing the interactions with photon field in a chiral invariant way.
The single heavy baryons have been studied
experimentally (e.g., Refs.~\cite{Aaltonen:2011sf,Aaij:2012da,Aaltonen:2013tta}),
and theoretically based on chiral models (e.g., Refs.~\cite{Cho:1994vg,Jiang:2015xqa,Cheng:2015naa}), 
quark models (e.g., Refs.~\cite{Ivanov:1999bk,Yoshida:2015tia,Hosaka:2016fnf,Nagahiro:2016nsx,Arifi:2017sac,Wang:2017kfr,Gandhi:2018lez}),
the sum rule (e.g., Refs.~\cite{Chen:2015kpa,Mao:2015gya,Chen:2017sci}),
the Regge theory (e.g., Ref.~\cite{Thakkar:2016dna}),
lattice simulations (e.g., Refs.~\cite{Bahtiyar:2016dom,Bahtiyar:2018vub}),
molecule models (e.g., Refs.~\cite{Guo:2017jvc,Lu:2014ina}) 
(See for a review, e.g., Ref.~\cite{Chen:2016spr} and references therein.).
In this paper, we make comparisons of our predictions with those in chiral effective models~\cite{Cho:1994vg,Jiang:2015xqa}, quark models~\cite{Ivanov:1999bk,Thakkar:2016dna,Wang:2017kfr}, and lattice simulations \cite{Bahtiyar:2016dom,Bahtiyar:2018vub}.

This paper is organized as follows:
We construct an effective Lagrangian in section~\ref{sec:Lag}.
Sections~\ref{sec:masses} and \ref{sec:2pi} are 
devoted to study the masses and
the hadronic decays of SHBs.
We also study the radiative decays of SHBs in section~\ref{sec:rad}.
Finally, we give a summary and discussions in section~\ref{sec:summary}.

\section{Effective Lagrangian}
\label{sec:Lag}

In this section, we construct an effective
model of single heavy baryons (SHBs) by extending the two-flavor model provided in the previous work~\cite{Kawakami:2018olq}, to three-flavor case.

We introduce a set of fields, $S_Q^\mu$ ($Q = b,c$), for SHBs in which the light-quark cloud carries the spin $1$ and belongs to $({\mathbf 3},{\mathbf 3})$ representation under $\mbox{SU}(3)_{L}\times \mbox{SU}(3)_{R}$ symmetry. The field  transforms as
\begin{equation}
	S_Q^\mu \stackrel{\mathrm{Ch.}}{\to} g_R S_Q^\mu g_L^T, \quad (Q = c, \ b)\ ,
\end{equation}
where $g_{L,R} \in \mbox{SU}(3)_{L,R}$.
When the chiral symmetry is spontaneously broken, $S_Q^\mu$ is divided into two parts. One is for the positive parity SHBs belonging to ${\mathbf 6}$ representation under $\mbox{SU}(3)_{\rm flavor}$ symmetry,  $\hat{B}_Q^{6 \mu}$, and another for the negative parity SHBs to $\bar{\mathbf 3}$, $\hat{B}_Q^{\bar{3} \mu}$:
\begin{equation}
	S_Q^\mu = \hat{B}_Q^{6 \mu} + \hat{B}_Q^{\bar{3} \mu}\ .
\end{equation}
We would like to stress that $\hat{B}_Q^{6 \mu}$ and $\hat{B}_Q^{\bar{3} \mu}$ are chiral partners to each other in the present model.
The physical states are embedded as
\begin{align}
	\hat{B}_Q^{6 \mu} & = 
	\begin{pmatrix} \Sigma_Q^{I=1 \mu} & \frac{1}{\sqrt{2}} \Sigma_Q^{I=0 \mu} & \frac{1}{\sqrt{2}} \Xi_Q^{\prime I=\frac{1}{2} \mu} \\
	\frac{1}{\sqrt{2}} \Sigma_Q^{I=0 \mu} & \Sigma_Q^{I=-1 \mu} & \frac{1}{\sqrt{2}}\Xi_Q^{\prime I=-\frac{1}{2} \mu} \\
	\frac{1}{\sqrt{2}} \Xi_Q^{\prime I=\frac{1}{2} \mu} & \frac{1}{\sqrt{2}} \Xi_Q^{\prime I=-\frac{1}{2} \mu} & \Omega_Q^\mu
	\end{pmatrix}\ , 
\notag\\
\hat{B}_Q^{\bar{3} \mu} & = \frac{1}{\sqrt{2}}
	\begin{pmatrix} 0 & \Lambda_{Q1}^\mu & \Xi_{Q1}^{I=\frac{1}{2} \mu} \\
	- \Lambda_{Q1}^\mu & 0 & \Xi_{Q1}^{I=-\frac{1}{2} \mu} \\
	- \Xi_{Q1}^{I=\frac{1}{2} \mu} & - \Xi_{Q1}^{I=-\frac{1}{2} \mu} & 0
	\end{pmatrix}\ .
\end{align}
These $B_Q^{6 \mu}$ and $B_Q^{\bar{3} \mu}$ are decomposed into spin-3/2 baryon fields and spin-1/2 fields as
\begin{align}
B_Q^{6 \mu} & = B_Q^{6 \ast \mu} - \frac{1}{\sqrt{3}} (\gamma^\mu + v^\mu)\gamma_5 B_Q^6\ , \notag\\
B_Q^{\bar{3} \mu} & = B_Q^{\bar{3} \ast \mu} - \frac{1}{\sqrt{3}} (\gamma^\mu + v^\mu) \gamma_5 B_Q^{\bar{3}}\ ,
\end{align}
where $B_Q^{6 \ast \mu}$ and $B_Q^{\bar{3} \ast \mu}$ denote the spin-3/2 baryon fields, and $B_Q^{6}$ and $B_Q^{\bar{3}}$ the spin-1/2 fields, respectively. We note that the parity transformation of the $S_Q^\mu$ field is given by 
\begin{equation}
	S_Q^\mu \stackrel{\mathrm{P}}{\to} - \gamma^0 S_{Q \mu}^T\ ,
\end{equation}
where $^T$ denotes the transposition of the $3 \times 3$ matrix in the light-quark flavor space, and that the Dirac conjugate is defined as 
\begin{equation}
	\bar{S}_Q^\mu = S_Q^{\mu \dagger} \gamma^0\ .
\end{equation}

We introduce a $3 \times 3$ matrix field $M$ for scalar and pseudoscalar mesons made from a light quark and a light anti-quark, which belongs to the $({\mathbf 3},\bar{\mathbf 3})$ representation under the chiral $\mbox{SU}(3)_L \times \mbox{SU}(3)_R$ symmetry. The transformation properties of $M$ under the chiral symmetry and the parity are given by 
\begin{align}
	M& \stackrel{\mathrm{Ch.}}{\to} g_L M g_R^\dagger\ , \\
	M& \stackrel{\mathrm{P}}{\to} M^\dagger\ .
\end{align}
We assume that the potential terms for $M$ in the model are constructed in such a way that the $M$ has a vacuum expectation value (VEV) which breaks the chiral symmetry spontaneously:
\begin{equation}
\langle M \rangle = \begin{pmatrix} f_\pi & 0 & 0 \\ 0 & f_\pi & 0 \\ 0 & 0 & \sigma_s \end{pmatrix} \ ,\label{vev}
\end{equation}
where $f_\pi$ is the pion decay constant and $\sigma_s$ is written as  $\sigma_s = 2f_K - f_\pi$ with the Kaon decay constant $f_K$.~\footnote{Here we adopt the normalization of $f_\pi = 92.4\,$MeV and $f_K = 1.197 f_\pi$}
In the following, for studying the decays of the SHBs with emitting 
pions,
we parameterize the field $M$ as 
\begin{equation}
M = \xi \, \begin{pmatrix} f_\pi & 0 & 0 \\ 0 & f_\pi & 0 \\ 0 & 0 & \sigma_s \end{pmatrix} \, \xi
\end{equation}
where
\begin{equation}
\xi = e^{ i \pi /f_\pi  } \ , 
\end{equation}
with $\pi$ being the $3 \times 3$ matrix field including 
pions
as
\begin{equation}
	\pi = \frac{1}{2}
\begin{pmatrix} \pi^0 & \sqrt{2} \pi^+ & 0 \\
 \sqrt{2} \pi^- & - \pi^0  & 0 \\
 0 & 0 & 0 \\ 	\end{pmatrix} \ .
\end{equation}

In addition, we introduce two fields, one belonging to $(\bar{\mathbf 3},{\mathbf 1})$ representation under $\mbox{SU}(3)_L \times \mbox{SU}(3)_R$ symmetry and another to $({\mathbf 1},\bar{\mathbf 3})$ representation.
It is convenient to use anti-symmetric $3 \times3$ matrix fields which transform as
\begin{equation}
	S_{QLL} \stackrel{\mathrm{Ch.}}{\to} g_L S_{QLL} g_L^T, \quad S_{QRR} \stackrel{\mathrm{Ch.}}{\to} g_R S_{QRR} g_R^T\ ,
\end{equation}
where $S_{QLL}$ and $S_{QRR}$ denote the fields of $(\bar{\mathbf 3},{\mathbf 1})$ and $({\mathbf 1},\bar{\mathbf 3})$ representations, respectively.
They are related to each others by parity transformation as
\begin{equation}
	S_{QLL} \stackrel{\mathrm{P}}{\to} - \gamma^0 S_{QRR}\ .
\end{equation} 
We introduce the parity eigenstates as
\begin{equation}
	S_{QLL} = \hat{A}^{\bar{3}}_Q - \hat{C}^{\bar{3}}_Q, \quad S_{QRR} = \hat{A}^{\bar{3}}_Q + \hat{C}^{\bar{3}}_Q\ ,
\end{equation}
where $A^{\bar{3}}_Q$ and $C^{\bar{3}}_Q$ carry the negative and positive parities, respectively.
They 
include the flavor anti-symmetric fields as
\begin{equation}
	\hat{A}^{\bar{3}}_Q =
	\begin{pmatrix} 0 & \Lambda_{Q2}^\mu & \Xi_{Q2}^{I=\frac{1}{2} \mu} \\
	- \Lambda_{Q2}^\mu & 0 & \Xi_{Q2}^{I=-\frac{1}{2} \mu} \\
	- \Xi_{Q2}^{I=\frac{1}{2} \mu} & - \Xi_{Q2}^{I=-\frac{1}{2} \mu} & 0
	\end{pmatrix}\ ,
\end{equation}
\begin{equation}
	\hat{C}^{\bar{3}}_Q =
	\begin{pmatrix} 0 & \Lambda_Q^\mu & \Xi_Q^{I=\frac{1}{2} \mu} \\
	- \Lambda_Q^\mu & 0 & \Xi_Q^{I=-\frac{1}{2} \mu} \\
	- \Xi_Q^{I=\frac{1}{2} \mu} & - \Xi_Q^{I=-\frac{1}{2} \mu} & 0
	\end{pmatrix}\ .
\end{equation}
These $A^{\bar{3}}_Q$ and $C^{\bar{3}}_Q$ express spin-1/2 fields respectively. Since the particles which are expressed by $A^{\bar{3}}_Q$ are still undiscovered, we neglect $A^{\bar{3}}_Q$ in the following discussion.

Now, let us write down an effective Lagrangian including the baryon fields $S_Q^\mu$, $S_{QLL}$, and $S_{QRR}$ together with the meson field $M$, based on the heavy-quark spin-flavor symmetry and the chiral symmetry. We do not consider the terms including more than square of $M$ field or more than two derivatives. A possible Lagrangian is given by
\begin{widetext}
\begin{align}
	\mathcal{L}_{Q}=&-{\rm{tr}}\bar{S}_{Q}^{\mu}\left(v\cdot iD-\Delta \right)S_{Q\mu}
+ \bar{S}_{QLL}\left(v\cdot iD\right)S_{QLL} + \bar{S}_{QRR}\left(v\cdot iD\right)S_{QRR} \notag\\
	&+\frac{g_{1}}{2f_{\pi}}{\rm{tr}}\left(\bar{S}_{Q}^{\mu}M^{\dagger}MS_{Q\mu}+\bar{S}_{Q\mu}^{T}MM^{\dagger}S_{Q}^{\mu T}\right)\notag\\
	&-\frac{g_{2}}{2f_{\pi}}{\rm{tr}}\bar{S}_{Q}^{\mu}M^{\dagger}S_{Q\mu}^{T}M^{T}
	-\frac{g_{2}^{v}}{2m_{\Lambda_{Q}}}{\rm{tr}}\bar{S}_{Q}^{\mu}M^{\dagger}S_{Q\mu}^{T}M^{T}\notag\\
	&+\frac{\kappa_1}{4f_\pi}{\rm{tr}}\left(\bar{S}_Q^\mu\mathcal{M}^\dagger MS_{Q\mu}+\bar{S}_Q^\mu M^\dagger \mathcal{M}S_{Q\mu}+\bar{S}_{Q\mu}^T\mathcal{M}M^\dagger S_Q^{\mu T}+\bar{S}_{Q\mu}^T M\mathcal{M}^\dagger S_Q^{\mu T}\right)\notag\\
	&-\frac{\kappa_2}{2f_\pi}{\rm{tr}}\left(\bar{S}_Q^\mu\mathcal{M}^\dagger S_{Q\mu}^T M^T+\bar{S}_Q^\mu M^\dagger S_{Q\mu}^T \mathcal{M}^T\right)\notag\\
	&-i\frac{h_{1}^{I}-ih_{1}^{R}}{4f_{\pi}^{2}}{\rm{tr}}\left(\bar{S}_{Q}^{\mu}M^{\dagger}v\cdot\partial MS_{Q\mu}+\bar{S}_{Q}^{\mu T}Mv\cdot\partial M^{\dagger}S_{Q\mu}^{T}\right)
	\notag\\
	&
	-i\frac{-h_{1}^{I}-ih_{1}^{R}}{4f_{\pi}^{2}}{\rm{tr}}\left(\bar{S}_{Q}^{\mu}v\cdot\partial M^{\dagger}MS_{Q\mu}+\bar{S}_{Q}^{\mu T}v\cdot\partial MM^{\dagger}S_{Q\mu}^{T}\right)\notag\\
	&+\frac{h_{2}}{2f_{\pi}^{2}}{\rm{tr}}\left(\bar{S}_{Q}^{\mu}v\cdot\partial M^{\dagger}S_{Q\mu}^{T}M^{T}+\bar{S}_{Q}^{\mu T}v\cdot\partial MS_{Q\mu}M^{*}\right) \notag \\
	&- \frac{g_3}{2 f_\pi} {\rm{tr}} \left( \bar{S}_{QLL} \partial^\mu M S_{Q\mu} + \bar{S}_Q^\mu \partial_\mu M^\dagger S_{QLL} + \bar{S}_{QRR} \partial^\mu M^\dagger S_{Q\mu}^T + \bar{S}_Q^{\mu T} \partial_\mu M S_{QRR}\right) \ ,
\end{align}
\end{widetext}
where $m_{\Lambda_{Q}}$ ($Q=c,b$) are the masses of $\Lambda_{c}(2286)$ and $\Lambda_{b}$ in the ground state,
$\Delta$ provides the difference between the chiral invariant mass of 
$S_Q^\mu$ and that of $S_{QLL}$ and $S_{QRR}$.
$g_i\ (i=1,2,3)$, $g_2^v$, $\kappa_i\ (i=1,2)$,  $h_1^I$, $h_1^R$, and $h_2$ are dimensionless coupling constants.
We note that we include $g_2^v$-term to incorporate
the heavy-flavor violation needed for explaining the mass differences of charm and bottom sectors (See Ref.~\cite{Kawakami:2018olq}.).
Although we can add heavy-quark flavor violation terms corresponding to $g_{1}$-term, such contributions are absorbed into the definition of $\Delta$.
We expect that heavy-quark flavor violating contributions to terms other than $g_2^v$ term are small. Since thresholds of $B_Q^{6 \ast} \to B_Q^6 \pi$ are not open, the related terms are not included here.
We note that the above chiral partner may not be necessarily a three-quark state but can be also a molecular state such as the one in Ref.~\cite{Lu:2014ina}.

\section{Masses and 
one-pion decays}
\label{sec:masses}

In this section, we determine the coupling constants $g_{2}$, $g_{2}^{v}$, and $\kappa_2$ from masses of relevant SHBs, and $g_{3}$ from $\Sigma_{c}^{(*)}\to\Lambda_{c}\pi$ decays. Then we make predictions of the one-pion decay widths of other members of the flavor $\bm{6}$ representation.

When the chiral symmetry is spotaneously broken, the light meson field $M$ acquires its vacuum expectation value as in Eq.~(\ref{vev}). Then the masses of the particles included in the model are expressed as 
\begin{widetext}
\begin{align}
	M(\Sigma_Q) &= M_{\Lambda_Q} + \Delta + g_1 f_\pi - \frac{g_2^Q}{2} f_\pi + \bar{\kappa}_1 - \bar{\kappa}_2 \\
	M(\Xi'_Q) &= M_{\Lambda_Q} + \Delta + g_1 \frac{f_\pi^2 + \sigma_s^2}{2 f_\pi} - \frac{g_2^Q}{2} \sigma_s + \bar{\kappa}_1 \frac{f_\pi + \sigma_s \frac{m_s}{\bar{m}}}{2 f_\pi} - \bar{\kappa}_2 \frac{f_\pi \frac{m_s}{\bar{m}} + \sigma_s}{2 f_\pi} \\ 
	M(\Omega_Q) &= M_{\Lambda_Q} + \Delta + g_1 \frac{\sigma_s^2}{f_\pi} - \frac{g_2^Q}{2} \frac{\sigma_s^2}{f_\pi} + \bar{\kappa}_1 \frac{m_s}{\bar{m}} \frac{\sigma_s}{f_\pi} - \bar{\kappa}_2 \frac{m_s}{\bar{m}} \frac{\sigma_s}{f_\pi} \\
	M(\Lambda_{Q1}) &= M_{\Lambda_Q} + \Delta + g_1 f_\pi + \frac{g_2^Q}{2} f_\pi + \bar{\kappa}_1 + \bar{\kappa}_2 \\ 
	M(\Xi_{Q1}) &= M_{\Lambda_Q} + \Delta + g_1 \frac{f_\pi^2 + \sigma_s^2}{2 f_\pi} + \frac{g_2^Q}{2} \sigma_s + \bar{\kappa}_1 \frac{f_\pi + \sigma_s \frac{m_s}{\bar{m}}}{2 f_\pi} + \bar{\kappa}_2 \frac{f_\pi \frac{m_s}{\bar{m}} + \sigma_s}{2 f_\pi}\ . 
\end{align}
\end{widetext}
where 
$\bar{\kappa}_i = \kappa_i \bar{m}$, and $g_2^Q$ is defined as
\begin{equation}
	g_2^Q = g_2 + g_2^v \frac{f_\pi}{m_{\Lambda_Q}}\ .
\end{equation}
We determine 
the fraction of strange quark mass $m_s$ and up or down quark mass $\bar{m}$ from the masses of the pion and kaon as $m_s/\bar{m} = 25.9$ using 
\begin{equation}
\frac{m_K^2}{m_\pi^2} = \frac{m_s + \bar{m} }{ 2 \bar{m} } \ .
\end{equation}
In the present analysis, we 
assign the following physical states to the flavor $\bm{\bar{3}}$ 
representation:
\begin{align}
& \left(\Lambda_{c1},\ \Lambda_{c1}^{*}\right)=\left(\Lambda_{c}(2595;\ J^{P}=1/2^{-})\, \Lambda_{c}(2625;\ 3/2^{-})\right) \ , \notag\\
& \left(\Xi_{c1},\ \Xi_{c1}^{*}\right)=\left(\Xi_{c}(2790;\ J^{P}=1/2^{-}),\ \Lambda_{c}(2815;\ 3/2^{-})\right)
\end{align}
and 
are the chiral partner to the flavor 
$\bm{6}$ representation:
\begin{align}
& \left(\Sigma_{c},\ \Sigma_{c}^{*}\right)=\left(\Sigma_{c}(2455;\ 1/2^{+}),\ \Sigma_{c}(2520;\ 3/2^{+})\right)\ ,\notag\\
& \left(\Xi^\prime_{c},\ \Xi_{c}^{\prime*}\right)=\left(\Xi_{c}^\prime(1/2^{+}),\ \Xi_{c}^\prime(3/2^{+})\right)\ , \notag\\
& \left(\Omega_{c},\ \Omega_{c}^{*}\right)=\left(\Omega_{c}(1/2^{+}),\ \Omega_{c}(2770;\ 3/2^{+})\right) \ .
\end{align}
In the  bottom sector, $\bm{\bar{3}}$ includes
\begin{align}
& \left(\Lambda_{b1},\ \Lambda_{b1}^{*}\right)=\left(\Lambda_{c}(5912;\ J^{P}=1/2^{-}),\ \Lambda_{c}(5920;\ 3/2^{-})\right)\ , \notag\\
& \left(\Xi_{b1},\ \Xi_{b1}^{*}\right)=\left(\Xi_{b}(1/2^-),\ \Xi_{b}(3/2^-)\right) \ , 
\end{align}
and $\bm{6}$ includes
\begin{align}
& \left(\Sigma_{b},\ \Sigma_{b}^{*}\right)=\left(\Sigma_{b}(1/2^{+}),\ \Sigma_{b}(3/2^{+})\right)\ ,\notag\\
& \left(\Xi^\prime_{b},\ \Xi_{b}^{\prime*}\right)=\left(\Xi_{b}^\prime(5935;\ 1/2^{+}),\ \Xi_{c}(5945,\ 5955; 3/2^{+})\right)\ , \notag\\
& \left(\Omega_{b},\ \Omega_{b}^{*}\right)=\left(\Omega_{b}(1/2^{+}),\ \Omega_{b}(3/2^{+})\right)\ .
\end{align} 
We list experimental data of their masses and full decay widths~\cite{Tanabashi:2018oca} in Table~\ref{exp. table}.
\begin{table}[h]
	\caption{Experimental data of masses and decay widths of heavy baryons included in the present analysis.}
	\begin{center}
		\begin{tabular}{ccccc} \hline \hline
		 particle & $J^{P}$ & mass[MeV] & full width[MeV]\\\hline
		 $\Lambda_{c}$ & $1/2^{+}$ & $2286.46\pm0.14$ & no strong decays\\\hline
		 $\Xi_c^+$ & $1/2^+$ & $2467.87 \pm 0.30$ & no strong decays \\
		 $\Xi_c^0$ & $1/2^+$ & $2470.87^{+0.28}_{-0.31}$ & no strong decays \\ \hline
		 $\Sigma_{c}^{++}(2455)$ & $1/2^{+}$ & $2453.97\pm0.14$ & $1.89^{+0.09}_{-0.18}$\\
		 $\Sigma_{c}^{+}(2455)$ & $1/2^{+}$ & $2452.9\pm0.4$ & $<4.6$\\
		 $\Sigma_{c}^{0}(2455)$ & $1/2^{+}$ & $2453.75\pm0.14$ & $1.83^{+0.11}_{-0.19}$\\\hline
		 $\Sigma_{c}^{++}(2520)$ & $3/2^{+}$ & $2518.41^{+0.21}_{-0.19}$ & $14.78^{+0.30}_{-0.40}$\\
		 $\Sigma_{c}^{+}(2520)$ & $3/2^{+}$ & $2517.5\pm1.3$ & $<17$\\
		 $\Sigma_{c}^{0}(2520)$ & $3/2^{+}$ & $2518.48\pm0.20$ & $15.3^{+0.4}_{-0.5}$\\\hline
		 $\Xi_c^{\prime +}$ & $1/2^+$ & $2577.4 \pm 1.2$ & no strong decays \\
		 $\Xi_c^{\prime 0}$ & $1/2^+$ & $2578.8 \pm 0.5$ & nostrong decays \\ \hline
		 $\Xi_c^+(2645)$ & $3/2^+$ & $2645.53 \pm 0.31$ & $2.14 \pm 0.19$ \\
		 $\Xi_c^0(2645)$ & $3/2^+$ & $2646.32 \pm 0.31$ & $2.35 \pm 0.18 \pm 0.13$ \\ \hline
		 $\Omega_c$ & $1/2^+$ & $2695.2 \pm 1.7$ & no strong decays \\
		 $\Omega_c(2770)$ & $3/2^+$ & $2765.9 \pm 2.0$ & no strong decays \\ \hline
		 $\Lambda_{c}(2595)$ & $1/2^{-}$ & $2595.25\pm0.28$ & $2.59\pm0.30\pm0.47$\\
		 $\Lambda_{c}(2625)$ & $3/2^{-}$ & $2628.11\pm0.19$ & $<0.97$\\\hline
		 $\Xi_c^+(2790)$ & $1/2^-$ & $2792.0 \pm 0.5$ & $8.9 \pm 0.6 \pm 0.8$ \\
		 $\Xi_c^0(2790)$ & $1/2^-$ & $2792.8 \pm 1.2$ & $10.0 \pm 0.7 \pm 0.8$ \\ \hline
		 $\Xi_c^+(2815)$ & $3/2^-$ & $2816.67 \pm 0.31$ & $2.43 \pm 0.20 \pm 0.17$ \\
		 $\Xi_c^0(2815)$ & $3/2^-$ & $2820.22 \pm 0.32$ & $2.54 \pm 0.18 \pm 0.14$ \\ \hline
		 $\Lambda_{b}$ & $1/2^{+}$ & $5619.58\pm0.17$ & no strong decays\\\hline
		 $\Xi_b^0$ & $1/2^+$ & $5791.9 \pm 0.5$ & no strong decays \\
		 $\Xi_b^-$ & $1/2^+$ & $5794.5 \pm 1.4$ & no strong decays \\ \hline
		 $\Sigma_{b}^{+}$ & $1/2^{+}$ & $5811.3^{+0.9}_{-0.8}\pm1.7$ & $9.7^{+3.8}_{-2.8}\ ^{+1.2}_{-1.1}$\\
		 $\Sigma_{b}^{0}$ & $1/2^{+}$ & $\cdots$ & $\cdots$ \\
		 $\Sigma_{b}^{-}$ & $1/2^{+}$ & $5815.5^{+0.6}_{-0.5}\pm1.7$ & $4.9^{+3.1}_{-2.1}\pm1.1$\\\hline
		 $\Sigma_{b}^{*+}$ & $3/2^{+}$ & $5832.1\pm0.7\ ^{+1.7}_{-1.8}$ & $11.5^{+2.7}_{-2.2}\ ^{+1.0}_{-1.5}$\\
		 $\Sigma_{b}^{*0}$ & $3/2^{+}$ & $\cdots$ & $\cdots$ \\
		 $\Sigma_{b}^{*-}$ & $3/2^{+}$ & $5835.1\pm0.6\ ^{+1.7}_{-1.8}$ & $7.5^{+2.2}_{-1.8}\ ^{+0.9}_{-1.4}$\\\hline
		 $\Xi_b^{\prime 0}$ & $1/2^+$ & $\cdots$ & $\cdots$ \\
		 $\Xi_b^{\prime -}(5935)$ & $1/2^+$ & $5935.02 \pm 0.02 \pm 0.05$ & $<0.08$ \\ \hline
		 $\Xi_b^0(5945)$ & $3/2^+$ & $5949.8 \pm 1.4$ & $0.90 \pm 0.16 \pm 0.08$ \\
		 $\Xi_b^-(5955)$ & $3/2^+$ & $5955.33 \pm 0.12 \pm 0.05$ & $1.65 \pm 0.31 \pm 0.10$ \\ \hline
		 $\Omega_b$ & $1/2^+$ & $6046.1 \pm 1.7$ & no strong decays \\
		 $\Omega_b^\ast$ & $3/2^+$ & $\cdots$ & $\cdots$ \\ \hline
		 $\Lambda_{b}(5912)$ & $1/2^{-}$ & $5912.18\pm0.13\pm0.17$ & $<0.66$\\
		 $\Lambda_{b}(5920)$ & $3/2^{-}$ & $5919.90\pm0.19$ & $<0.63$\\\hline
		 $\Xi_{b1}^0$ & $1/2^-$ & $\cdots$ & $\cdots$ \\
		 $\Xi_{b1}^-$ & $1/2^-$ & $\cdots$ & $\cdots$ \\ \hline
		 $\Xi_{b1}^0$ & $3/2^-$ & $\cdots$ & $\cdots$ \\
		 $\Xi_{b1}^-$ & $3/2^-$ & $\cdots$ & $\cdots$ \\ \hline
		\end{tabular}
		\label{exp. table}
	\end{center}
\end{table} 

Here, we cannot determine the values of $\Delta$, $g_1$, and $\kappa_1$, separately. Instead, we introduce
\begin{align}
	\bar{\Delta} & = \Delta + g_1 f_\pi + \bar{\kappa}_1 \notag \\
	\Delta_s & = \Delta + g_1 \frac{f_\pi^2 + \sigma_s^2}{2 f_\pi} + \bar{\kappa}_1 \frac{f_\pi + \sigma_s \frac{m_s}{\bar{m}}}{2 f_\pi} \notag \\
	\Delta_\Omega & = \Delta + g_1 \frac{\sigma_s^2}{f_\pi} + \bar{\kappa}_1 \frac{m_s}{\bar{m}} \frac{\sigma_s}{f_\pi} \ ,
\end{align}
to rewrite mass formulas as
\begin{align}
	M(\Sigma_Q) &= M_{\Lambda_Q} +\bar{\Delta} - \frac{g_2^Q}{2} f_\pi - \bar{\kappa}_2 \ , \notag \\ 
	M(\Xi'_Q) &= M_{\Lambda_Q} +\Delta_s - \frac{g_2^Q}{2} \sigma_s - \bar{\kappa}_2 \frac{f_\pi \frac{m_s}{\bar{m}} + \sigma_s}{2 f_\pi} \ , \notag \\
	M(\Omega_Q) &= M_{\Lambda_Q} + \Delta_\Omega - \frac{g_2^Q}{2} \frac{\sigma_s^2}{f_\pi} - \bar{\kappa}_2 \frac{m_s}{\bar{m}} \frac{\sigma_s}{f_\pi}\ , \notag \\ 
	M(\Lambda_{Q1}) &= M_{\Lambda_Q} +\bar{\Delta} + \frac{g_2^Q}{2} f_\pi+ \bar{\kappa}_2  \ , \notag \\ 
	M(\Xi_{Q1}) &= M_{\Lambda_Q} +\Delta_s + \frac{g_2^Q}{2} \sigma_s + \bar{\kappa}_2 \frac{f_\pi \frac{m_s}{\bar{m}} + \sigma_s}{2 f_\pi} \ . 
\end{align}
We estimate the values of mass parameters and coupling constants in charm sector from experimental data in a way explained in Ref.~\cite{Kawakami:2018olq}:
We
calculate the spin-averaged mass of SHBs in a heavy-quark multiplet with including errors to include the masses of members belonging to the multiplet as shown in Table~\ref{spin averaged mass}.
\begin{table}[h]
\caption{
Spin averaged masses and widths used as inputs to determine the model parameters.
}
\begin{center}
		\begin{tabular}{cc} \hline \hline
		 input & value (MeV) \\ \hline
		 $M\left(\Lambda_c\right)$ & $2286.46$ \\
		 $M\left(\Sigma_c^{(\ast)}\right)$ & $2496.6^{+21.5}_{-43.6}$ \\
		 $M\left(\Xi_c^{(\ast)}\right)$ & $2623.3^{+22.6}_{-45.2}$ \\
		 $M\left(\Omega_c^{(\ast)}\right)$ & $2742.3^{+23.6}_{-47.1}$ \\
		 $M\left(\Lambda_{c1}^{(\ast)}\right)$ & $2617.16^{+10.95}_{-21.91}$ \\
		 $M\left(\Xi_{c1}^{(\ast)}\right)$ & $2809.8^{+8.7}_{-17.4}$ \\
		 $\Gamma\left(\Sigma_c^{(\ast)} \to \Lambda_c \pi\right)$ & $10.6^{+4.9}_{-9.0}$ \\ \hline
		 $M\left(\Lambda_b\right)$ & $5619.58$ \\
		 $M \left(\Lambda_{b1}^{(\ast)}\right) - M \left(\Sigma_b^{(\ast)}\right)$ & $90.5^{+8.3}_{-4.2}$ \\
\hline
\end{tabular}
	\end{center}
\label{spin averaged mass}
\end{table} 

To include the heavy quark flavor symmetry violation, we determined the value of $g_2^b$ from the mass difference between spin-averaged masses of 
$\Lambda_{b1}^{(\ast)}$ and $\Sigma_b^{(\ast)}$.
In addition, we use the weighted average of $\Sigma_{c}^{++}\to\Lambda_{c}^{+}\pi^{+},\ \Sigma_{c}^{0}\to\Lambda_{c}^{+}\pi^{-},\ \Sigma_{c}^{\ast ++}\to\Lambda_{c}^{+}\pi^{+}$ and $\Sigma_{c}^{\ast 0}\to\Lambda_{c}^{+}\pi^{-}$  to determine the coupling constant $g_{3}$ as done in Ref.~\cite{Kawakami:2018olq}.
We show the estimated values of model parameters in Table~\ref{tab: g2 g3}.
\begin{table}[h]
\caption{Estimated values of model paramters}
\begin{center}
\begin{tabular}{cc} \hline \hline
		 parameter & value \\ \hline
		 $\bar{\Delta}$ & $270^{+17}_{-34}$\,MeV  \\
		 $\Delta_s$ & $430^{+16}_{-31}$\,MeV  \\
		 $ \Delta_\Omega $ & $600^{+13}_{-27}$\,MeV \\
		 $g_2^c$ & $1.28^{+0.20}_{-0.11}$  \\
		 $g_2^b$ & $0.980^{+0.090}_{-0.046}$ \\
		 $\bar{\kappa}_2$ & $0.807^{+0.47}_{-0.23}$ \\
		 $g_3$ & $0.688^{+0.013}_{-0.025}$  \\ \hline
		\end{tabular}
	\end{center}
\label{tab: g2 g3}
\end{table} 

Using the estimated value of $g_3$, we predict the decay widths of $\Sigma_Q^{(\ast)} \to \Lambda_Q \pi$ and $\Xi_Q^{\prime (\ast)} \to \Xi_Q \pi$ as shown in Table~\ref{decay of SigmaQ}. These predictions are consistent with experimental data because light flavor symmetry violation and heavy quark symmetry violation are small for the $g_3$-term.
\begin{table}[H]
\caption{Decay widths $\Sigma_{Q}^{(*)}\to\Lambda_{Q}\pi$ predicted in our model. }
	\begin{center}
		\begin{tabular}{cccccc} \hline \hline
		 decay modes & our model [MeV] & expt. [MeV] \\\hline
		 $\Sigma_{c}^{++}\to\Lambda_{c}^{+}\pi^{+}$ & $1.96^{+0.07}_{-0.14}$ & $1.89^{+0.09}_{-0.18}$\\
		 $\Sigma_{c}^{+}\to\Lambda_{c}^{+}\pi^{0}$ & $2.28^{+0.09}_{-0.17}$ & $<4.6$\\
		 $\Sigma_{c}^{0}\to\Lambda_{c}^{+}\pi^{-}$ & $1.94^{+0.07}_{-0.14}$ & $1.83^{+0.11}_{-0.19}$\\\hline
		 $\Sigma_{c}^{*++}\to\Lambda_{c}^{+}\pi^{+}$ & $14.7^{+0.6}_{-1.1}$ & $14.78^{+0.30}_{-0.40}$\\
		 $\Sigma_{c}^{*+}\to\Lambda_{c}^{+}\pi^{0}$ & $15.3^{+0.6}_{-1.1}$ & $<17$\\
		 $\Sigma_{c}^{*0}\to\Lambda_{c}^{-}\pi^{0}$ & $14.7^{+0.6}_{-1.1}$ & $15.3^{+0.4}_{-0.5}$\\\hline
		 $\Sigma_{b}^{+}\to\Lambda_{b}^{0}\pi^{+}$ & $6.14^{+0.23}_{-0.45}$ & $9.7^{+3.8}_{-2.8}\ ^{+1.2}_{-1.1}$\\
		 $\Sigma_{b}^{0}\to\Lambda_{b}^{0}\pi^{0}$ & $7.27^{+0.27}_{-0.53}$ & $\cdots$ \\
		 $\Sigma_{b}^{-}\to\Lambda_{b}^{0}\pi^{-}$ & $7.02^{+0.27}_{-0.51}$ & $4.9^{+3.1}_{-2.1}\pm1.1$\\\hline
		 $\Sigma_{b}^{*+}\to\Lambda_{b}^{0}\pi^{+}$ & $11.0^{+0.4}_{-0.8}$ & $11.5^{+2.7}_{-2.2}\ ^{+1.0}_{-1.5}$\\
		 $\Sigma_{b}^{*0}\to\Lambda_{b}^{0}\pi^{0}$ & $12.3^{+0.5}_{-0.9}$ & $\cdots$ \\
		 $\Sigma_{b}^{*-}\to\Lambda_{b}^{0}\pi^{-}$ & $11.9^{+0.4}_{-0.9}$ & $7.5^{+2.2}_{-1.8}\ ^{+0.9}_{-1.4}$\\\hline
		 $\Xi_c^{\prime +} \to \Xi_c^+ \pi^0$ & $\multirow{2}{*}{$\cdots$}$ & \multirow{2}{*}{no strong decays} \\
		 $\Xi_c^{\prime +} \to \Xi_c^0 \pi^+$ & $$ &  \\ \hline
		 $\Xi_c^{\prime 0} \to \Xi_c^+ \pi^-$ & $\multirow{2}{*}{$\cdots$}$ & \multirow{2}{*}{no strong decays} \\
		 $\Xi_c^{\prime 0} \to \Xi_c^0 \pi^0$ & $$ &  \\ \hline
		 $\Xi_c^{\prime \ast +} \to \Xi_c \pi$ & $2.39^{+0.09}_{-0.17}$ & $2.14 \pm 0.19$ \\
		 $\Xi_c^{\prime \ast +} \to \Xi_c^+ \pi^0$ & $0.953^{+0.036}_{-0.069}$ & $\cdots$ \\
		 $\Xi_c^{\prime \ast +} \to \Xi_c^0 \pi^+$ & $1.44^{+0.05}_{-0.10}$ & $\cdots$ \\ \hline
		 $\Xi_c^{\prime \ast 0} \to \Xi_c \pi$ & $2.57^{+0.10}_{-0.19}$ & $2.35 \pm 0.18 \pm 0.13$ \\
		 $\Xi_c^{\prime \ast 0} \to \Xi_c^0 \pi^0$ & $0.873^{+0.033}_{-0.063}$ & $\cdots$ \\
		 $\Xi_c^{\prime \ast 0} \to \Xi_c^+ \pi^-$ & $1.70^{+0.06}_{-0.12}$ & $\cdots$ \\ \hline
		 $\Xi_b^{\prime 0} \to \Xi_b \pi$ & $0.0806^{+0.0030}_{-0.0059}$ & $\cdots$ \\
		 $\Xi_b^{\prime 0} \to \Xi_b^0 \pi^0$ & $0.0746^{+0.0028}_{-0.0054}$ & $\cdots$ \\
		 $\Xi_b^{\prime 0} \to \Xi_b^- \pi^+$ & $0.00601^{+0.00023}_{-0.00044}$ & $\cdots$ \\ \hline
		 $\Xi_b^{\prime -} \to \Xi_b \pi$ & $0.0853^{+0.0032}_{-0.0062}$ & $<0.08$ \\
		 $\Xi_b^{\prime -} \to \Xi_b^- \pi^0$ & $0.0413^{+0.0016}_{-0.0030}$ & $\cdots$ \\
		 $\Xi_b^{\prime -} \to \Xi_b^0 \pi^-$ & $0.0440^{+0.0017}_{-0.0032}$ & $\cdots$ \\ \hline
		 $\Xi_b^{\prime \ast 0} \to \Xi_b \pi$ & $0.813^{+0.031}_{-0.059}$ & $0.90 \pm 0.16 \pm 0.08$ \\
		 $\Xi_b^{\prime \ast 0} \to \Xi_b^0 \pi^0$ & $0.378^{+0.014}_{-0.027}$ & $\cdots$ \\
		 $\Xi_b^{\prime \ast 0} \to \Xi_b^- \pi^+$ & $0.435^{+0.016}_{-0.032}$ & $\cdots$ \\ \hline
		 $\Xi_b^{\prime \ast -} \to \Xi_b \pi$ & $1.30^{+0.05}_{-0.09}$ & $1.65 \pm 0.31 \pm 0.10$ \\
		 $\Xi_b^{\prime \ast -} \to \Xi_b^- \pi^0$ & $0.459^{+ 0.017}_{-0.033}$ & $\cdots$ \\
		 $\Xi_b^{\prime \ast -} \to \Xi_b^0 \pi^-$ & $0.843^{+0.032}_{-0.061}$ & $\cdots$ \\ \hline
		\end{tabular}
		\label{decay of SigmaQ}
	\end{center}
\end{table} 

We can estimate the masses of bottom baryons included in our model using the parameters in Table~\ref{tab: g2 g3}. 
In the present analysis, we assume heavy quark spin symmetry, so that we predict the spin-averaged masses which 
are shown in Table~\ref{bottom mass}.
Here, we show the result in Ref.~\cite{Thakkar:2016dna, Chen:2018orb} and experimental values for comparison.  
We note that, 
in Table~\ref{bottom mass}, we just put the minimum and maximum values predicted for the members in a multiplet in Ref.~\cite{Thakkar:2016dna}.
This table shows that our predictions are consistent with those in Ref.~\cite{Thakkar:2016dna, Chen:2018orb}.

\begin{table}[h]
\caption{Predicted values of the spin-averaged masses of bottom baryons. For comparison we list the spin-averages of experimentally observed masses and the predicted values in Ref.~\cite{Thakkar:2016dna, Chen:2018orb}. }
	\begin{center}
		\begin{tabular}{ccccc} \hline \hline
		 particle & our model & \cite{Thakkar:2016dna} & \cite{Chen:2018orb} & expt. \\
		 & & & & (spin averaged) \\ \hline
		 $\Sigma_b^{(\ast)}$ & $5843^{+20}_{-37}$ & $5811 - 5835$ & $\cdots$ & $5826.9$ \\
		 $\Xi_b^{\prime(\ast)}$ & $5975^{+18}_{-37}$ & $\cdots$ & $\cdots$ & $5946.7$ \\
		 $\Omega_b^{(\ast)}$ & $6102^{+15}_{-36}$ & $6048 - 6086$ & $\cdots$ & $6046.1$ (spin-1/2) \\
		 $\Lambda_{b1}^{(\ast)}$ & $5936^{+20}_{-36}$ & $5980 - 6000$ & $\cdots$ & $5917.33$ \\
		 $\Xi_{b1}^{(\ast)}$ & $6124^{+20}_{-34}$ & $6129 - 6151$ & $6096$, $6102$ & $\cdots$ \\ \hline
		\end{tabular}
	\end{center}
\label{bottom mass}
\end{table} 

We can see that our predictions for $\Sigma_b^{(\ast)}$, $\Xi_b^{\prime(\ast)}$ and $\Lambda_{b1}^{(\ast)}$ are consistent with the spin-averaged masses of experimentally observed masses. 
For $\Omega_b$, only  the mass of the spin-$1/2$ member is known experimentally.  
Although our prediction of the spin-average is slightly larger than the observed mass of the spin-$1/2$ member, we expect that the spin-$3/2$ member is slightly heavier which makes the spin-averaged larger and consistent with our prediction. 
Future experimental observation of spin-$3/2$ member as well as $\Xi_{b1}^{(\ast)}$ will be a test of the present model. 
We note that $\Xi_{b1}^{(\ast)}$ in the present analysis
are unlikely to make a multiplet including $\Xi_b(6227)$ reported in Ref.~\cite{Aaij:2018yqz},
since the predicted mass of $\Xi_{b1}^{(\ast)}$ is about 100 MeV smaller than the observed mass of $\Xi_b(6227)$. 

\section{Pion decays of single heavy baryons with negative parity}
\label{sec:2pi}

In this section, we consider decays of $B_Q^{\bar{3} (\ast)}$, the negative parity excited SHBs belonging to the flavor $\bm{3}$ representation.
The main modes of $\Lambda_{Q1}^{(\ast)}$ are three body decay, $\Lambda_{Q1} \to \Lambda_Q \pi \pi$ because $\Lambda_{Q1}^{(\ast)} \to \Sigma_Q^{(\ast)} \pi$ decay thresholds are closed in most cases. In the decays of $\Xi_{c1}^{(\ast)}$, the decay thresholds of $\Xi_c^{(\ast)} \to \Xi_c^{\prime (\ast)} \pi$ are completely open, so the main mode is the two body decay. 

\begin{figure}[H]
	\centering
	\includegraphics[width=8cm]{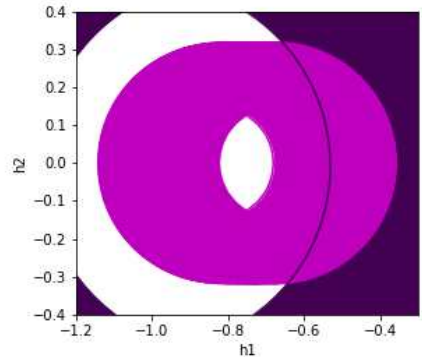}
	\caption{Allowed range of $h_{1}^{I}$ and $h_{2}$.}
	\label{h1h2}
\end{figure}

In Ref.~\cite{Kawakami:2018olq}, we used the two-pion decay width of $\Lambda_c(2595)$ to determine the values of derivative coupling constants, $h_1^I$ and $h_2$.
Here, we also include the decay widths of $\Xi_c (2790)$ and $\Xi_c (2815)$.  
There exists violation of the heavy quark spin symmetry between the decay widths of $\Xi_c (2790)$ and $\Xi_c (2815)$.
Instead of treating this violation precisely, we include the violation as systematic errors of the model. 
Therefore, we use values of a decay width of $\Lambda_c(2595)$ and, a spin averaged decay width between $\Xi_c (2790)$ and $\Xi_c (2815)$ as inputs to determine  $h_1^I$ and $h_2$.
The region colored by dark purple in Fig.~\ref{h1h2}
shows the allowed values of $h_1^I$ and $h_2$ determined from the decay width of $\Lambda_c (2595)$ where 
the errors of $g_2^c$, $g_3$ and the total width with $\Lambda(2595)$
are taken into account.
The region by light purple are obtained from the spin averaged width of $\Xi_c (2790)$ and $\Xi_c (2815)$ with the errors of model parameters included.

In the following analysis,
we use the values of $h_1^I$ and $h_2$ in the overlapped region of two colors in Fig.~\ref{h1h2} to make predictions of the decay widths of excited SHBs with negative pariry.
We show the results of total decay widths in Table~\ref{total width}, where we list predictions by the quark model in Refs.~\cite{Chen:2018orb, Wang:2017kfr} for comparison.
\begin{widetext}
\begin{table}[H]
\caption{Predicted widths of excited SHBs. We used the spin and isospin averaged value of the decay widths of $\Xi_c^+(2790)$, $\Xi_c^0(2790)$, $\Xi_c^+ (2815)$ and $\Xi_c^0 (2815)$ in addition to the decay width of $\Lambda_c(2595)$ as inputs.}
\label{3bd}
\begin{center}
\begin{tabular}{cccccc} \hline \hline
initial & mode & Our model & \cite{Chen:2018orb} & \cite{Wang:2017kfr} & expt.\\ 
 & & [MeV] & [MeV] & [MeV] & [MeV] \\
\hline
$\Lambda_{c}(2595)$ & $\Lambda_c\pi^+\pi^-$ & $0.562$-$1.09$  &  \\
 & $\Lambda_c\pi^0\pi^0$ & $1.23$-$2.31$ & \\
 & sum & $1.82$-$3.36$ (input) & $\cdots$ & $\cdots$ & $2.59\pm0.30\pm0.47$ \\
\hline
$\Lambda_{c}(2625)$ & $\Lambda_c\pi^+\pi^-$ & $0.0618$-$0.507$ & \\
& $\Lambda_c\pi^0\pi^0$ & $0.0431$-$0.226$ & \\
& sum & $0.106$-$0.733$ & $\cdots$ & $\cdots$ & $<0.97$ \\
\hline
$\Lambda_{b}(5912)$ & $\Lambda_b\pi^+\pi^-$ & $(0.67$-$4.4)\times10^{-3}$ &  \\
 & $\Lambda_b\pi^0\pi^0$ & $(1.4$-$6.0)\times10^{-3}$ &  \\
 & sum & $(2.1$-$10)\times10^{-3}$  & $\cdots$ & $\cdots$ & $<0.66$ \\
\hline
$\Lambda_{b}(5920)$ & $\Lambda_b\pi^+\pi^-$ & $(0.75$-$13)\times10^{-3}$ & \\
 & $\Lambda_b\pi^0\pi^0$ & $(2.2$-$12)\times10^{-3}$  & \\
 & sum & $(3.0$-$25)\times10^{-3}$ & $\cdots$ & $\cdots$ & $<0.63$ \\
\hline
$\Xi_c^+ (2790)$ & $\Xi_c^{\prime +} \pi^0$ & $$ \\
 & $\Xi_c^{\prime 0} \pi^+$ & $$ \\
 & sum & input & $\cdots$ & $3.61$ & $8.9 \pm 0.6 \pm 0.8$ \\ \hline
$\Xi_c^0 (2790)$ & $\Xi_c^{\prime 0} \pi^0$ & $$ \\
 & $\Xi_c^{\prime +} \pi^-$ & $$ \\
 & sum & input & $\cdots$ & $3.61$ & $10.0 \pm 0.7 \pm 0.8$ \\ \hline
$\Xi_c^+ (2815)$ & $\Xi_c^+ (2645) \pi^0$ & $$ \\
 & $\Xi_c^0 (2645) \pi^+$ & $$ \\
 & sum & input & $\cdots$ & $1.80$ & $2.43 \pm 0.20 \pm 0.17$ \\ \hline
$\Xi_c^0 (2815)$ & $\Xi_c^0 (2645) \pi^0$ & $$ \\
 & $\Xi_c^+ (2645) \pi^-$ & $$ \\
 & sum & input & $\cdots$ & $1.80$ & $2.54 \pm 0.18 \pm 0.14$ \\ \hline
$\Xi_{b1}^0$ & $\Xi_b^{\prime 0} \pi^0$ & $0.0140 - 5.18$ \\
 & $\Xi_b^{\prime -} \pi^+$ & $0.0275 - 10.1$ \\
 & sum & $0.0415 - 15.3$ & $4.2$ & $2.84$ & $\cdots$ \\ \hline
$\Xi_{b1}^-$ & $\Xi_b^{\prime -} \pi^0$ & $0.0140 - 5.18$ \\
 & $\Xi_b^{\prime 0} \pi^-$ & $0.0275 - 10.1$ \\
 & sum & $0.0415 - 15.3$ & $4.2$ & $2.84$ & $\cdots$ \\ \hline
$\Xi_{b1}^{\ast 0}$ & $\Xi_b^{\ast 0} \pi^0$ & $0.109 - 4.31$ \\
 & $\Xi_b^{\ast -} \pi^+$ & $< 7.21$ \\
 & sum & $0.326 - 11.5$ & $2.9$ & $2.88$ & $\cdots$ \\ \hline
$\Xi_{b1}^{\ast -}$ & $\Xi_b^{\ast -} \pi^0$ & $0.109 - 4.31$ \\
 & $\Xi_b^{\ast 0} \pi^-$ & $< 7.21$ \\
 & sum & $0.326 - 11.5$ & $2.9$ & $2.88$ & $\cdots$ \\ \hline
\end{tabular}		
\end{center}
\label{total width}
\end{table} 
\end{widetext}

We note that we use the predicted masses of $\Xi_b$ with negative parity shown in Table~\ref{bottom mass} with including their errors.
So the predicted decay widths take a wide range of values, which includes predictions in Refs.~\cite{Chen:2018orb, Wang:2017kfr}.  In particular, since
the minimum value shown in Table~\ref{bottom mass} is very close to the threshold of the relevant decays, the minimum values of the predictions of one-pion decays of $\Xi_{b1}^{(\ast)}$ in Table~\ref{total width} are very small, and 
three-body decays such as $\Xi_{b1} \to \Xi_b \pi \pi$ become dominant. 
Here, we study the contributions of possible intermediate states of three-body decays and show the results in Table~\ref{Xi 3bd}, where 
we set the parameters as $g_2^b = 0.980, \ \bar{\kappa}_2 = 0.807, \ g_3 = 0.688, \ h_1^I = -0.40$, and $h_2 = 0$, 
and set the masses of $\Xi_{b1}^\ast$ to be their minimum of the predicted values shown in Table~\ref{bottom mass}, 
and $\Xi_b^{\prime \ast}$ to be the central mass values in Table~\ref{exp. table}.
\begin{table}[htbp]
	\caption{Estimated values of the decay widths of bottom baryons.}
	\begin{center}
		\begin{tabular}{cccc} \hline \hline
		 initial & decay & intermediate & width\\
		 state & mode & state & [keV] \\		 \hline
		 $\Xi_{b1}^{\ast 0}$ & $\Xi_b^0 \pi^0 \pi^0$ & non-resonant & $0.688$ \\
		 & & $\Xi_b^{\prime \ast 0}$ & $534$ \\
		 & & NR \& $\Xi_b^{\prime \ast 0}$ & $2.97$ \\ \hline
		 & $\Xi_b^0 \pi^+ \pi^-$ & NR & $0.432$ \\
		 & & $\Xi_b^{\prime \ast -}$ & $8.67$ \\
		 & & NR \& $\Xi_b^{\prime \ast -}$ & $2.59$ \\ \hline
		 & $\Xi_b^- \pi^+ \pi^0$ & $\Xi_b^{\prime \ast 0}$ & $2.44 \times 10^3$ \\
		 & & $\Xi_b^{\prime \ast -}$ & $41.1$ \\
		 & & $\Xi_b^{\prime \ast 0}$ \& $\Xi_b^{\prime \ast -}$ & $2.06$ \\ \hline \hline
		 $\Xi_{b1}^{\ast -}$ & $\Xi_b^0 \pi^0 \pi^0$ & NR & $0.510$ \\
		 & & $\Xi_b^{\prime \ast -}$ & $53.6$ \\
		 & & NR \& $\Xi_b^{\prime \ast -}$ & $5.02$ \\ \hline
		 & $\Xi_b^- \pi^+ \pi^-$ & NR & $0.275$ \\
		 & & $\Xi_b^{\prime \ast 0}$ & $194$ \\
		 & & NR \& $\Xi_b^{\prime \ast 0}$ & $4.55$ \\ \hline
		 & $\Xi_b^0 \pi^- \pi^0$ & $\Xi_b^{\prime \ast -}$ & $386$ \\
		 & & $\Xi_b^{\prime \ast 0}$ & $1.40 \times 10^3$ \\
		 & & $\Xi_b^{\prime \ast -}$ \& $\Xi_b^{\prime \ast 0}$ & $5.39$ \\ \hline \hline
		\end{tabular}
	\end{center}
\label{Xi 3bd}
\end{table} 
We note that,
unlikely to the decays of $\Lambda_b(5912)$ and $\Lambda_b(5920)$,
the decays of $\Xi_{b1}^{\ast0}$ and $\Xi_{b1}^{\ast-}$ are not dominated by the non-resonant contribution.

\newpage
\section{Radiative decays}
\label{sec:rad}

In this section, we study radiative decays of the SHBs. The relevant Lagrangian is given by
\begin{align}
	\mathcal{L}_{\mathrm{rad}} & = \frac{r_1}{F} \mathrm{tr} \left(\bar{S}^\mu_Q Q_{\mathrm{light}} S_Q^\nu + \bar{S}_Q^{\mu T} Q_{\mathrm{light}} S_Q^{\nu T} \right) F_{\mu \nu} \notag \\
	& + \frac{r_2}{F} \mathrm{tr} \left(\bar{S}_Q^\mu Q_{\mathrm{light}} S_Q^\nu - \bar{S}_Q^{\mu T} Q_{\mathrm{light}} S_Q^{\nu T} \right) \tilde{F}_{\mu \nu} \notag \\
	& + \frac{r_3}{F^2} \mathrm{tr} \left(\bar{S}_{QLL} M S_Q^\mu Q_\mathrm{light} v^\nu + \bar{S}_{QRR} M^\dagger S_Q^\mu Q_\mathrm{light} v^\nu\right) F_{\mu \nu} \notag \\
	& + \mathrm{h.c.} \notag \\
	& + \frac{r_4}{F^2}\mathrm{tr} \left(\bar{S}_{QLL} M S_Q^\mu Q_\mathrm{light} v^\nu + \bar{S}_{QRR} M^\dagger S_Q^\mu Q_\mathrm{light} v^\nu\right) \tilde{F}_{\mu \nu} \notag \\
	& + \mathrm{h.c.} \ ,
\label{rad}
\end{align}
where $F_{\mu \nu}$ is the field strength of the photon and $\tilde{F}_{\mu \nu}$ is its dual tensor: $\tilde{F}_{\mu \nu} = (1/2) \epsilon_{\mu \nu \rho \sigma} F^{\rho \sigma}$, $r_i \ (i = 1,...,4)$ are dimensionless constants, and $F$ is a constant with dimension one. In this analysis, we take $F = 350$ MeV following Ref.~\cite{Cho:1994vg}. We note that the values of the constants $r_i$ are of order one based on quark models~\cite{Cho:1994vg}.

Let us first study the electromagnetic intramultiplet transitions governed by the $r_1$-term in Eq.~(\ref{rad}). Let $B^\ast$ denotes the decaying baryon with spin-3/2 ($B^\ast = \Lambda_{Q1}^\ast,\ \Xi_{Q1}^\ast,\ \Sigma_Q^\ast,\ \Xi_Q^{\prime \ast},\ \Omega_Q^\ast$), and $B$, the daughter baryon with spin-1/2 ($\Lambda_{Q1},\ \Xi_{Q1},\ \Sigma_Q,\ \Xi_Q^\prime,\ \Omega_Q$). Then the radiative decay width is given by
\begin{equation}
\Gamma_{B^{*}\to B\gamma} = C_{{B^{*}B\gamma}}^{2}\,
\frac{ 16 \alpha r_1^2 }{9F^2} 
\frac{m_{B}}{m_{B^{*}}}E_{\gamma}^{3}
\end{equation}
where $\alpha$ is the electromagnetic fine structure constant, $E_\gamma$ is the photon energy and $C_{B^\ast B \gamma}$ is the Clebsh-Gordon constants given by
\begin{align}
	C_{\Sigma_{c}^{*++}\Sigma_{c}^{++}\gamma}=C_{\Sigma_{b}^{*+}\Sigma_{b}^{+}\gamma}&=\frac{2}{3}, \notag\\
	C_{\Sigma_{c}^{*+}\Sigma_{c}^{+}\gamma}=C_{\Sigma_{b}^{*0}\Sigma_{b}^{0}\gamma}&=\frac{1}{6}, \notag \\
	C_{\Sigma_{c}^{*0}\Sigma_{c}^{0}\gamma}=C_{\Sigma_{b}^{*-}\Sigma_{b}^{-}\gamma}&=-\frac{1}{3},\notag \\
	C_{\Xi_c^{\ast +} \Xi_c^+ \gamma} = C_{\Xi_b^{\ast 0} \Xi_b^0 \gamma} & = \frac{1}{6}, \notag \\
	C_{\Xi_c^{\ast 0} \Xi_c^0 \gamma} = C_{\Xi_b^{\ast -} \Xi_b^- \gamma} & = -\frac{1}{3}, \notag \\
	C_{\Omega_c^{\ast 0} \Omega_c^0 \gamma} = C_{\Omega_b^{\ast -} \Omega_b^- \gamma} & = -\frac{1}{3}, \notag \\
	C_{\Lambda_{c1}^{*+}\Lambda_{c1}^{+}\gamma} = C_{\Lambda_{b}^{*0}\Lambda_{b}^{0}\gamma} & = -\frac{1}{6}, \notag \\
	C_{\Xi_{c1}^{\ast +} \Xi_{c1}^+ \gamma} = C_{\Xi_{b1}^{\ast 0} \Xi_{b1}^0 \gamma} & = -\frac{1}{6}, \notag \\
	C_{\Xi_{c1}^{\ast 0} \Xi_{c1}^0 \gamma} = C_{\Xi_{b1}^{\ast -} \Xi_{b1}^- \gamma} & = \frac{1}{3}\ .
\end{align}

Here we would like to stress that the radiative decay widths of positive parity SHBs ($\Sigma_Q^{\ast}$, $\Xi_Q^{\ast}$, $\Omega_Q^{\ast}$) and those of negative parity SHBs ($\Lambda_{Q1}^\ast$, $\Xi_{Q1}^\ast$) are determined by just one coupling constant $r_1$, reflecting the chiral partner structure.
We think that checking the relation among these radiative decays will be one of the crucial test of the chiral partner structure.
In Table~\ref{r1 charm} and \ref{r1 bottom}, we show our predictions 
comparing with those
in Ref.~\cite{Jiang:2015xqa,Bahtiyar:2018vub}.
In radiative decays, the chiral loop considered in Ref.~\cite{Jiang:2015xqa} might have contribution.
However, our predictions are consistent with those in Ref.~\cite{Jiang:2015xqa} for $r_1 \sim 1$,
which implies the contribution from the chiral loop is small in this radiative decay.
On the other hand, our results are consistent with the lattice results if $r_1 \sim 0.2$.

We note that 
$\Omega_Q^\ast$ does not have any strong decays and the main mode must be $\Omega_Q^\ast \to \Omega_Q \gamma$. We expect the coupling constant $r_1$ will be determined by the decay of $\Omega_Q^\ast$ in future experiments, and other radiative decay widths related to $r_1$-type interaction will be estimated.
\begin{table}[h]
\caption{Predicted widths of radiative decays
between heavy quark multiplets of charm baryons.
We also show the predictions in Ref.~\cite{Jiang:2015xqa, Bahtiyar:2018vub} for comparison.}
	\begin{center}
		\begin{tabular}{cccc} \hline \hline
		 decay mode & predicted width & \cite{Jiang:2015xqa} & \cite{Bahtiyar:2018vub} \\ 
		  & [keV] & [keV] & [keV] \\
		 \hline
		 $\Sigma_c^{\ast ++} \to \Sigma_c^{++} \gamma$ & $11.8 r_1^2$ & $11.6$ & $\cdots$ \\
		 $\Sigma_c^{\ast +} \to \Sigma_c^+ \gamma$ & $0.743 r_1^2$ & $0.85$ & $\cdots$ \\
		 $\Sigma_c^{\ast 0} \to \Sigma_c^0 \gamma$ & $2.99 r_1^2$ & $2.92$ & $\cdots$ \\ \hline
		 $\Xi_c^{\prime \ast +} \to \Xi_c^{\prime +} \gamma$ & $0.872 r_1^2$ & $1.10$ & $\cdots$ \\
		 $\Xi_c^{\prime \ast 0} \to \Xi_c^{\prime 0} \gamma$ & $3.40 r_1^2$ & $3.83$ & $\cdots$ \\ \hline
		 $\Omega_c^{\ast 0} \to \Omega_c^0 \gamma$ & $3.90 r_1^2$ & $4.82$ & $0.096(14)$ \\ \hline
		 $\Lambda_{c1}^{\ast +} \to \Lambda_{c1}^+ \gamma$ & $0.131 r_1^2$ & $\cdots$ & $\cdots$ \\ \hline
		 $\Xi_{c1}^{\ast +} \to \Xi_{c1}^+ \gamma$ & $0.0432 r_1^2$ & $\cdots$ & $\cdots$ \\
		 $\Xi_{c1}^{\ast 0} \to \Xi_{c1}^0 \gamma$ & $0.237 r_1^2$ & $\cdots$ & $\cdots$ \\ \hline
		\end{tabular}
	\end{center}
\label{r1 charm}
\end{table} 

\begin{table}[h]
\caption{
Predicted widths of radiative decays
between heavy quark multiplets of bottom baryons.
We also show the predictions in Ref.~\cite{Jiang:2015xqa} for comparison.
}
	\begin{center}
		\begin{tabular}{ccc} \hline \hline
		 decay mode & predicted width & \cite{Jiang:2015xqa} \\
		  & [keV] & [keV] \\
		 \hline
		 $\Sigma_b^{\ast +} \to \Sigma_b^+ \gamma$ & $0.420 r_1^2$ & $0.60$ \\
		 $\Sigma_b^{\ast 0} \to \Sigma_b^0 \gamma$ & $0.0240 r_1^2$ & $0.05$ \\
		 $\Sigma_b^{\ast -} \to \Sigma_b^- \gamma$ & $0.0879 r_1^2$ & $0.08$ \\ \hline
		 
		 $\Xi_b^{\prime \ast 0} \to \Xi_b^{\prime 0} \gamma$ & $0.00944 r_1^2$ & $\cdots$ \\
		 $\Xi_b^{\prime \ast -} \to \Xi_b^{\prime -} \gamma$ & $0.0977 r_1^2$ & $\cdots$ \\ \hline
		 
		 $\Omega_b^{\ast -} \to \Omega_b^- \gamma$ & $0.0920 - 4.07 r_1^2$ & $\cdots$ \\ \hline
		 $\Lambda_{b1}^{\ast 0} \to \Lambda_{b1}^0 \gamma$ & $0.00135 r_1^2$ & $\cdots$ \\ \hline
		 $\Xi_{b1}^{\ast 0} \to \Xi_{b1}^0 \gamma$ & $< 0.453 r_1^2$ & $\cdots$ \\
		 $\Xi_{b1}^{\ast -} \to \Xi_{b1}^- \gamma$ & $< 1.81 r_1^2$ & $\cdots$ \\ \hline
		 
		\end{tabular}
	\end{center}
\label{r1 bottom}
\end{table} 

We next study the radiative decays between the SHBs with negative parity in the flavor $\bm{3}$ representations and the SHBs with positive parity in the flavor $\bm{6}$ representations,
which concern the $r_2$-term. The decay widths are expressed as
\begin{align}
	\Gamma_{\Lambda_{Q1}\to\Sigma_{Q}\gamma} = & \frac{ 16 \alpha r_2^2 }{9 F^2} 
\frac{m_{\Sigma_{Q}}}{m_{\Lambda_{Q1}}}E_{\gamma}^{3} 
\ , \notag\\
	\Gamma_{\Lambda_{Q1}\to\Sigma^{*}_Q\gamma} = & \frac{ 8 \alpha r_2^2}{9 F^2}
\frac{m_{\Sigma^{*}_Q}}{m_{\Lambda_{Q1}}}E_{\gamma}^{3} 
\ , \notag\\
	\Gamma_{\Lambda_{Q1}^{*}\to\Sigma_Q\gamma} = & \frac{4 r_2^2}{9 F^2}
\frac{m_{\Sigma_Q}}{m_{\Lambda_{Q1}^{*}}}E_{\gamma}^{3}
\ , \notag\\
	\Gamma_{\Lambda_{Q1}^{*}\to\Sigma^{*}_Q\gamma} = & \frac{20 \alpha r_2^2}{9 F^2}
\frac{m_{\Sigma^{*}_Q}}{m_{\Lambda_{Q1}^{*}}}E_{\gamma}^{3} \notag \\
	\Gamma_{\Xi_{Q1}^+ \to \Xi_{Q}^{\prime +} \gamma} = & \frac{ 16 \alpha r_2^2 }{9 F^2} 
\frac{m_{\Xi_{Q}^{\prime +}}}{m_{\Xi_{Q1}^+}}E_{\gamma}^{3} 
\ , \notag\\
	\Gamma_{\Xi_{Q1}^+ \to \Xi^{\prime \ast +}_Q \gamma} = & \frac{ 8 \alpha r_2^2}{9 F^2}
\frac{m_{\Xi^{\prime \ast +}_Q}}{m_{\Xi_{Q1}^+}}E_{\gamma}^{3} 
\ , \notag\\
	\Gamma_{\Xi_{Q1}^{\ast +} \to \Xi_Q^{\prime +} \gamma} = & \frac{4 r_2^2}{9 F^2}
\frac{m_{\Xi_Q^{\prime +}}}{m_{\Xi_{Q1}^{\ast +}}}E_{\gamma}^{3}
\ , \notag\\
	\Gamma_{\Xi_{Q1}^{\ast +} \to \Xi^{\prime \ast +}_Q\gamma} = & \frac{20 \alpha r_2^2}{9 F^2}
\frac{m_{\Xi^{\prime \ast +}_Q}}{m_{\Xi_{Q1}^{\ast +}}}E_{\gamma}^{3} \ .
\end{align}
In Table~\ref{r2 charm} and \ref{r2 bottom}, we show our predictions comparing with those in Ref.~\cite{Cho:1994vg,Ivanov:1999bk}.
Our results are consistent with those in Ref.~\cite{Cho:1994vg} when $r_2 \sim c_{RS}/\sqrt{2}$,
and with those in Ref.~\cite{Ivanov:1999bk} when $r_2 \sim 1/2$.

\begin{table}[h]
\caption{
Predicted widths of radiative decays
between negative parity charm baryons 
in the flavor $\bm{3}$ representations
and positive parity charm baryons in the flavor $\bm{6}$ representations.
We also show the predictions in Ref.~\cite{Cho:1994vg, Ivanov:1999bk}  for comparison.}
	\begin{center}
		\begin{tabular}{cccc} \hline \hline
		 decay mode & predicted width & \cite{Cho:1994vg} & \cite{Ivanov:1999bk} \\
		  & [keV] & [keV] & [keV] \\ \hline
		 $\Lambda_{c1}^+ \to \Sigma_c^+ \gamma$ & $250 r_2^2$ & $127 c_{RS}^2$ & $77 \pm 1$ \\
		 $\Lambda_{c1}^+ \to \Sigma_c^{\ast +} \gamma$ & $20.6 r_2^2$ & $6 c_{RS}^2$ & $6 \pm 0.1$ \\ \hline
		 $\Lambda_{c1}^{\ast +} \to \Sigma_c^+ \gamma$ & $120 r_2^2$ & $58 c_{RS}^2$ & $35 \pm 0.5$ \\
		 $\Lambda_{c1}^{\ast +} \to \Sigma_c^{\ast +} \gamma$ & $161 r_2^2$ & $54 c_{RS}^2$ & $46 \pm 0.6$ \\ \hline
		 $\Xi_{c1}^+ \to \Xi_c^{\prime +} \gamma$ & $859 r_2^2$ & $\cdots$ & $\cdots$ \\
		 $\Xi_{c1}^+ \to \Xi_c^{\prime \ast +} \gamma$ & $146 r_2^2$ & $\cdots$ & $\cdots$ \\ \hline
		 $\Xi_{c1}^{\ast +} \to \Xi_c^{\prime +} \gamma$ & $291 r_2^2$ & $\cdots$ & $\cdots$ \\
		 $\Xi_{c1}^{\ast +} \to \Xi_c^{\prime \ast +} \gamma$ & $568 r_2^2$ & $\cdots$ & $\cdots$ \\ \hline
		\end{tabular}
	\end{center}
\label{r2 charm}
\end{table} 

\begin{table}[h]
\caption{Predicted widths of radiative decays
between negative parity bottom baryons 
in the flavor $\bm{3}$ representations
and positive parity bottom baryons in the flavor $\bm{6}$ representations.
}
	\begin{center}
		\begin{tabular}{cc} \hline \hline
		 decay mode & predicted width \\
		  & [keV]  \\
		  \hline
		 $\Lambda_{b1}^0 \to \Sigma_b^0 \gamma$ & $97.9 r_2^2$ \\
		 $\Lambda_{b1}^0 \to \Sigma_b^{\ast 0} \gamma$ & $24.8 r_2^2$ \\ \hline
		 $\Lambda_{b1}^{\ast 0} \to \Sigma_b^0 \gamma$ & $30.6 r_2^2$ \\
		 $\Lambda_{b1}^{\ast 0} \to \Sigma_b^{\ast 0} \gamma$ & $82.0 r_2^2$ \\ \hline
		 $\Xi_{b1}^0 \to \Xi_b^{\prime 0} \gamma$ & $370 - 887 r_2^2$ \\
		 $\Xi_{b1}^0 \to \Xi_b^{\prime \ast 0} \gamma$ & $138 - 358 r_2^2$ \\ \hline
		 $\Xi_{b1}^{\ast 0} \to \Xi_b^{\prime 0} \gamma$ & $92.4 - 222 r_2^2$ \\
		 $\Xi_{b1}^{\ast 0} \to \Xi_b^{\prime \ast 0} \gamma$ & $344 - 895 r_2^2$ \\ \hline
		\end{tabular}
	\end{center}
\label{r2 bottom}
\end{table} 

The $r_3$-term generates the 
radiative decays between the negative parity SHBs in the flavor $\bm{3}$ representations and the positive parity SHBs in the flavor $\bm{3}$ representations,
the widths of which are expressed as 
\begin{align}
	\Gamma_{\Lambda_{Q1}^{(*)}\to\Lambda_Q\gamma} & =\frac{8 \alpha r_{3}^{2}f_{\pi}^{2}}{27 F^{4}}\frac{m_{\Lambda_Q}}{m_{\Lambda_{Q1}^{(*)}}}E_{\gamma}^{3} \ , \notag \\
	\Gamma_{\Xi_{Q1}^{(\ast) +} \to \Xi_Q^+ \gamma} & =\frac{8 \alpha r_{3}^{2}(f_{\pi} - 2 \sigma_s)^{2}}{27 F^{4}}\frac{m_{\Xi_Q^+}}{m_{\Xi_{Q1}^{(\ast) +}}}E_{\gamma}^{3} \ , \notag \\
	\Gamma_{\Xi_{Q1}^{(\ast) 0} \to \Xi_Q^0 \gamma} & =\frac{8 \alpha r_{3}^{2}(f_{\pi} + \sigma_s)^{2}}{27 F^{4}}\frac{m_{\Xi_Q^0}}{m_{\Xi_{Q1}^{(\ast) 0}}}E_{\gamma}^{3} \ .
\end{align}
In Tabel~\ref{r3 charm} and \ref{r3 bottom}, we show our predictions together with the ones in Ref.~\cite{Cho:1994vg,Ivanov:1999bk}.
We think that the
differences between our predictions and those in Ref.~\cite{Ivanov:1999bk} are from the value of $\sigma_s$:
we use $\sigma_s = 2 f_K - f_\pi$ while $\sigma_s = f_\pi$ is used in Ref.~\cite{Ivanov:1999bk}.

\begin{table}[h]
\caption{
Predicted widths of radiative decays
between negative parity charm baryons in the flavor $\bm{3}$ representations
and positive parity charm baryons in the flavor $\bm{3}$ representations. We also show the predictions
in Ref.~\cite{Cho:1994vg, Ivanov:1999bk}.
}
	\begin{center}
		\begin{tabular}{cccc} \hline \hline
		 decay mode & predicted width & \cite{Cho:1994vg} & \cite{Ivanov:1999bk} \\
		  & [keV] & [keV] & [keV] \\
		 \hline
		 $\Lambda_{c1} \to \Lambda_c \gamma$ & $25.9 r_3^2$ & $191 c_{RT}^s$ & $115 \pm 1$ \\
		 $\Lambda_{c1}^\ast \to \Lambda_c \gamma$ & $34.9 r_3^2$ & $253 c_{RT}^2$ & $151 \pm 2$ \\ \hline
		 $\Xi_{c1}^+ \to \Xi_c^+ \gamma$ & $98.9 r_3^2$ & $\cdots$ & $\cdots$ \\
		 $\Xi_{c1}^{\ast +} \to \Xi_c^+ \gamma$ & $121 r_3^2$ & $\cdots$ & $190 \pm 5$ \\ \hline
		 $\Xi_{c1}^0 \to \Xi_c^0 \gamma$ & $174 r_3^2$ & $\cdots$ & $\cdots$ \\
		 $\Xi_{c1}^{\ast 0} \to \Xi_c^0 \gamma$ & $217 r_3^2$ & $\cdots$ & $497 \pm 14$ \\ \hline
		 
		\end{tabular}
	\end{center}
\label{r3 charm}
\end{table} 
\begin{table}[h]
\caption{Predicted widths of radiative decays
between negative parity bottom baryons in the flavor $\bm{3}$ representations and positive parity bottom baryons in the flavor $\bm{3}$ representations.}
	\begin{center}
		\begin{tabular}{cc} \hline \hline
		 decay mode & predicted width \\
		 & [keV] \\ \hline 
		 $\Lambda_{b1} \to \Lambda_b \gamma$ & $27.2 r_3^2$ \\
		 $\Lambda_{b1}^\ast \to \Lambda_b \gamma$ & $29.3 r_3^2$ \\ \hline
		 $\Xi_{b1}^0 \to \Xi_b^0 \gamma$ & $92.0 - 148 r_3^2$ \\
		 $\Xi_{b1}^{\ast 0} \to \Xi_b^0 \gamma$ & $92.0 - 148 r_3^2$ \\ \hline
		 $\Xi_{b1}^- \to \Xi_b^- \gamma$ & $161 - 260 r_3^2$ \\
		 $\Xi_{b1}^{\ast -} \to \Xi_b^- \gamma$ & $161 - 260 r_3^2$ \\ \hline
		 
		\end{tabular}
	\end{center}
\label{r3 bottom}
\end{table} 

The widths of 
radiative decays between the positive parity SHBs in the flavor $\bm{6}$ representations and the positive parity SHBs in the flavor $\bm{3}$ representations
via the $r_4$-term are given by
\begin{align}
	\Gamma_{\Sigma^{(*)}_Q\to\Lambda_Q\gamma} & =\frac{8 \alpha r_{4}^{2}f_{\pi}^{2}}{3 F^{4}}\frac{m_{\Lambda_Q}}{m_{\Sigma^{(*)}_Q}}E_{\gamma}^{3}\ , \notag \\
	\Gamma_{\Xi^{\prime (\ast) +}_Q \to \Xi_Q^+ \gamma} & =\frac{8 \alpha r_{4}^{2}(f_{\pi} + 2 \sigma_s)^{2}}{27 F^{4}}\frac{m_{\Xi_Q^+}}{m_{\Xi^{\prime (\ast) +}_Q}}E_{\gamma}^{3}\ , \notag \\
	\Gamma_{\Xi^{\prime (\ast) 0}_Q \to \Xi_Q^0 \gamma} & =\frac{8 \alpha r_{4}^{2}(f_{\pi} - \sigma_s)^{2}}{27 F^{4}}\frac{m_{\Xi_Q^0}}{m_{\Xi^{\prime (\ast) 0}_Q}}E_{\gamma}^{3}\ ,
\end{align}
and the predicted values are shown in Table~\ref{r4 charm} and \ref{r4 bottom} with the ones in Ref.~\cite{Jiang:2015xqa,Bahtiyar:2016dom}.
Our results are consistent with the lattice results in Ref.~\cite{Bahtiyar:2016dom} if $r_4 \sim 0.1$.
On the other hand, comparison with the results in Ref.~\cite{Jiang:2015xqa} indicates that the chiral loop may be important.

\begin{table}[h]
\caption{Predicted widths of radiative decays
between positive parity
charm baryons in the flavor $\bm{6}$ representations
and positive parity charm baryons in the flavor $\bm{3}$ representations. We also show the predictions
in Ref.~\cite{Jiang:2015xqa, Bahtiyar:2016dom}. }
	\begin{center}
		\begin{tabular}{cccc} \hline \hline
		 decay mode & predicted width & \cite{Jiang:2015xqa} & \cite{Bahtiyar:2016dom} \\
		  & [keV] & [keV] & [keV] \\
		 \hline
		 $\Sigma_c^+ \to \Lambda_c^+ \gamma$ & $42.9 r_4^2$ & $164$ & $\cdots$ \\
		 $\Sigma_c^{\ast +} \to \Lambda_c^+ \gamma$ & $108 r_4^2$ & $893$ & $\cdots$ \\ \hline
		 $\Xi_c^{\prime +} \to \Xi_c^+ \gamma$ & $20.8 r_4^2$ & $54.3$ & $5.468(1.500)$ \\
		 $\Xi_c^{\prime \ast +} \to \Xi_c^+ \gamma$ & $83.3 r_4^2$ & $502$ & $\cdots$ \\ \hline
		 $\Xi_c^{\prime 0} \to \Xi_c^0 \gamma$ & $0.216 r_4^2$ & $0.02$ & $0.002(4)$ \\
		 $\Xi_c^{\prime \ast 0} \to \Xi_c^0 \gamma$ & $0.870 r_4^2$ & $0.36$ & $\cdots$ \\ \hline
		 
		\end{tabular}
	\end{center}
\label{r4 charm}
\end{table} 

\begin{table}[h]
\caption{Predicted widths of radiative decays
between positive parity
bottom baryons in the flavor $\bm{6}$ representations
and positive parity
bottom baryons in the flavor $\bm{3}$ representations.  
We also show the predictions
in Ref.~\cite{Jiang:2015xqa}.}
	\begin{center}
		\begin{tabular}{ccc} \hline \hline
		 decay mode & predicted width & \cite{Jiang:2015xqa} \\
		 & [keV] & [keV] \\ \hline
		 $\Sigma_b^0 \to \Lambda_b^0 \gamma$ & $74.1 r_4^2$ & $288$ \\
		 $\Sigma_b^{\ast 0} \to \Lambda_b^0 \gamma$ & $98.9 r_4^2$ & $435$ \\ \hline
		 $\Xi_b^{\prime 0} \to \Xi_b^0 \gamma$ & $48.7 r_4^2$ & $\cdots$ \\
		 $\Xi_b^{\prime \ast 0} \to \Xi_b^0 \gamma$ & $65.0 r_4^2$ & $136$ \\ \hline
		 $\Xi_b^{\prime -} \to \Xi_b^- \gamma$ & $0.499 r_4^2$ & $\cdots$ \\
		 $\Xi_b^{\prime \ast -} \to \Xi_b^- \gamma$ & $0.742 r_4^2$ & $1.87$ \\ \hline
		 
		\end{tabular}
	\end{center}
\label{r4 bottom}
\end{table}

\section{A summary and discussions}
\label{sec:summary}

We constructed an effective hadronic model regarding negative parity $\mathbf{3}$ representations as chiral partners to positive parity $\mathbf{6}$ representations, 
based on the chiral symmetry and heavy-quark spin-flavor symmetry. We determine the model parameters from the experimental data for relevant masses and decay widths of $\Sigma_c (2455, 1/2^+)$, $\Sigma_c (2520, 3/2^+)$, $\Lambda_c (2595, 1/2^-)$, $\Xi_c (2790, 1/2^-)$, and $\Xi_c (2815, 1/2^-)$. Then, we studied the decay widths of $\Lambda_c (2625)$, $\Lambda_b (5912)$, $\Lambda_b (5920)$, and 
negative parity $\Xi_b^{(\ast)}$ which have not been yet discovered in any experiments
We think that $\Xi_b^{(\ast)}$ here is unlikely to be $\Xi_b(6227)$ reported in Ref.~\cite{Aaij:2018yqz},
which may be explained as e.g., a molecule state in Ref.~\cite{Huang:2018bed}.
Using the model parameters, we predict the values for masses and decay widths of negative parity excited $\Xi_b$.

As shown in our previous work Ref.~\cite{Kawakami:2018olq}, the chiral partner structure is reflected in the direct decay processes in three-body decays of negative parity $\mathbf{3}$ representations.
Our results for the three-body decays of $\Xi_{b1}^{\ast} \to \Xi_b \pi \pi$ are dominated by 
the resonant decay modes unlikely to the decays of $\Lambda_c(2625)$, $\Lambda_b(5912)$ and $\Lambda_b(5920)$ shown in Ref.~\cite{Kawakami:2018olq}. 
However, the Dalitz analysis, which was performed in Ref.~\cite{Arifi:2018yhr} for the decays of $\Lambda_c$s and $\Lambda_b$s, 
may give an information of the direct decays of $\Xi$s. 
Therefore, we 
would like to stress that future investigations of detailed three-body decay processes of negative parity SHBs will provide some clues to understand the chiral partner structure.

We also studied the radiative decays of the SHBs included in the present model using the effective interaction Lagrangians in Eq.~(\ref{rad}). We showed that there is a relation among the radiative decay widths of positive parity SHBs ($\Sigma_Q^{\ast}$, $\Xi_Q^{\ast}$, $\Omega_Q^{\ast}$) and those of negative parity SHBs ($\Lambda_{Q1}^\ast$, $\Xi_{Q1}^\ast$), reflecting the chiral partner structure.
Since the masses of negative parity SHBs in the bottom sector are close to the threshold of hadronic decays, the radiative decay widths can be comparable with the strong decay widths depending on the precise values of the masses.  We summarize the decays of bottom SHBs with negative parity in Table~\ref{tab:summary}.
\newpage
\begin{table}[h]
\caption{
Pionic and radiative decays of bottom SHBs with negative parity.
} 
\label{tab:summary}
\begin{center}
\begin{tabular}{ccccc} \hline \hline
SHB & $J^{P}$ & decay & Our model & exp. \\
 & & modes  & [MeV] & [MeV] \\
\hline
$\Lambda_{b1}$ & $1/2^{-}$ & $\Lambda_{b}\pi^+\pi^-$ & $(0.67$-$4.4)\times10^{-3}$  & \multirow{5}{*}{$<0.66$} \\
 & & $\Lambda_{b}\pi^0\pi^0$ & $(1.4$-$6.0)\times10^{-3}$ & \\
 & & $\Sigma_b^0\gamma$ & $0.098\,r_2^2$ & \\
 & & $\Sigma_b^{\ast0}\gamma$ & $0.025\,r_2^2$ & \\
 & & $\Lambda_b\gamma$ & $0.027\,r_3^2$ & \\
\hline
$\Lambda_{b1}^\ast$ & $3/2^{-}$ & $\Lambda_{b}\pi^+\pi^-$ & $(0.75$-$13)\times10^{-3}$   & \multirow{6}{*}{$<0.63$} \\
 & & $\Lambda_{b}\pi^0\pi^0$ & $(2.2$-$12)\times10^{-3}$ & \\
 & & $\Lambda_{b1}\gamma$ & $0.0013 \, r_1^2 \times 10^{-3}$ & \\
 & & $\Sigma_b^0\gamma$ & $0.031\,r_2^2$ & \\
 & & $\Sigma_b^{\ast0}\gamma$ & $0.081\,r_2^2$ & \\
 & & $\Lambda_b\gamma$ & $0.029\,r_3^2$ & \\
\hline
$\Xi_{b1}^0$ & $1/2^-$ & $\Xi_b^\prime \pi$ & $0.0415 - 15.3$ & \multirow{4}{*}{$\cdots$} \\
 & & $\Xi_b^{\prime 0} \gamma$ & $0.370 - 0.887 r_2^2$ & \\
 & & $\Xi_b^{\prime \ast 0} \gamma$ & $0.138 - 0.358 r_2^2$ & \\
 & & $\Xi_b^0 \gamma$ & $0.0288 - 0.0464 r_3^2$ & \\ \hline
$\Xi_{b1}^-$ & $1/2^-$ & $\Xi_b^\prime \pi$ & $0.0415 - 15.3$ & \multirow{4}{*}{$\cdots$} \\
 & & $\Xi_b^{\prime -} \gamma$ & $\cdots$ & \\
 & & $\Xi_b^{\prime \ast -} \gamma$ & $\cdots$ & \\
 & & $\Xi_b^- \gamma$ & $0.112 - 0.182 r_3^2$ & \\ \hline
$\Xi_{b1}^{\ast 0}$ & $3/2^-$ & $\Xi_b^\prime \pi$ & $0.326 - 11.5$ & \multirow{5}{*}{$\cdots$} \\
 & & $\Xi_{b1}^0 \gamma$ & $< 4.53 r_1^2 \times 10^{-4}$ & \\
 & & $\Xi_b^{\prime 0} \gamma$ & $0.0924 - 0.222 r_2^2$ & \\
 & & $\Xi_b^{\prime \ast 0} \gamma$ & $0.344 - 0.895 r_2^2$ & \\
 & & $\Xi_b^0 \gamma$ & $0.0288 - 0.0464 r_3^2$ &\\ \hline
$\Xi_{b1}^{\ast -}$ & $3/2^-$ & $\Xi_b^\prime \pi$ & $0.326 - 11.5$ & \multirow{5}{*}{$\cdots$} \\
 & & $\Xi_{b1}^- \gamma$ & $< 1.81 r_1^2 \times 10^{-3}$ & \\
 & & $\Xi_b^{\prime -} \gamma$ & $\cdots$ & \\
 & & $\Xi_b^{\prime \ast -} \gamma$ & $\cdots$ & \\
 & & $\Xi_b^- \gamma$ & $0.112 - 0.182 r_3^2$ &\\ \hline
\end{tabular}
\end{center}
\end{table} 
We expect that experimental study of these radiative decays will provide a clue to understand the chiral partner structure. In addition, 
we predict the $\Omega_Q^{(\ast)} \to \Omega_Q \gamma$ decay which is the sole decay mode of $\Omega_Q^{(\ast)}$.
Experimental observation of this in future will be a check of the present framework based on the effective model respecting the chiral symmetry and the heavy-quark spin-flavor symmetry. In addition, we expect that the future lattice simulations for the 
radiative decay of negative parity SHBs 
also provide some clues to the chiral partner structure.

While we are writing this manuscript, we are informed
of Ref.~\cite{DCH}, in which the chiral partner structure of SHBs is 
studied based on the mirror assignment of parity doublet structure including 
three chiral representations, i.e., $(\bm{3},\bm{3})$, $(\bar{\bm{3}},\bm{1})+(\bm{1},\bar{\bm{3}})$
and $(\bm{6},\bm{1})+(\bm{1},\bm{6})$.

\subsection*{Acknowledgments}
We would like to thank Prof. Veljko Dmitra\v{s}inovi\'c, Prof. Hua-Xing Chen and Prof. Atsushi Hosaka
for their useful comments.

The work of M.H. is supported in part by 
JPSP KAKENHI
Grant Number 16K05345.

\end{document}